\documentclass[%
 reprint,
 amsmath,amssymb,
 aps,
]{revtex4-1}

\usepackage{graphicx}
\usepackage{dcolumn}
\usepackage{bm}
\usepackage{dsfont}
\usepackage{appendix}

\usepackage{color}
\usepackage{ulem}

\begin{document}

\preprint{APS/123-QED}

\title{Unimodular gauge fixing in general relativity revisited}

\author{Renata Jora
	$^{\it \bf a}$~\footnote[1]{Email:
		rjora@theory.nipne.ro}}
\email[ ]{rjora@theory.nipne.ro}

\affiliation{$^{\bf \it a}$ National Institute of Physics and Nuclear Engineering PO Box MG-6, Bucharest-Magurele, Romania}

\begin{abstract}

We study a slight variation of the unimodular gauge condition which we call "the derivative gauge condition". We show that at the classical level these conditions are completely equivalent up to a surface term. At the quantum level the derivative gauge condition can be easily implemented by analogy with gauge fixing in  a quantum gauge theory.  This condition might have advantages and disadvantages upon the context.

\end{abstract}
\maketitle

\section{Introduction}

Consider the action for general relativity:
\begin{eqnarray}
S=\int \sqrt{-g}d^4x\Bigg[\frac{1}{2k}R+{\cal L}_m\Bigg],
\label{actgenrel56774}
\end{eqnarray}
where $R$ is the curvature and ${\cal L}_m$ is the Lagrangian of matter. Then the Einstein equations can be derived readily by taking the functional derivative $\frac{\delta S}{\delta g^{\mu\nu}}$ with the result:
\begin{eqnarray}
R_{\mu\nu}-\frac{1}{2}g_{\mu\nu}R=\frac{8\pi G}{c^4}T_{\mu\nu},
\label{eisnt5665}
\end{eqnarray}
where $R_{\mu\nu}$ is the Ricci tensor, $R$ is the curvature and $T_{\mu\nu}$ is the matter  energy momentum tensor.

Equation (\ref{eisnt5665}) contains  second order partial differential equations for the metric tensor.  In general a second rank tensor has $16$ components and if this tensor is symmetric the number of independent component reduces to $10$.  The Bianchi identities ($\partial_{\mu}G^{\mu\nu}=0$ where $G_{\mu\nu}=R_{\mu\nu}-\frac{1}{2}g_{\mu\nu}R$) further reduces the number of independent components to $6$.

In quantum field theories the redundancy of the gauge degrees of freedom is resolved by finding a suitable gauge fixing condition. One or another gauge fixing condition may be suitable in a different theoretical or phenomenological context. In the case of electrodynamics only there are a few most common gauge fixing conditions as for example: Lorentz gauge $\partial_{\mu}A^{\mu}=0$, temporal gauge $A_0=0$ or radiation or transverse gauge $\partial_kA^k=0$ where $k$ indicates the spatial components).

In general relativity gauge fixing implies $4$ constraints that should be applied to the metric. The analogue for the Lorentz gauge condition for electromagnetism is the so called harmonic or De Donder gauge condition $g^{\mu\nu}\Gamma^{\rho}_{\mu\nu}=0$. The harmonic gauge condition may be written in different ways and stems from the harmonicity condition for the space time variables which amounts to:
\begin{eqnarray}
\Box x^{\mu}=0
\label{harmon567746}
\end{eqnarray}
where $\Box=\nabla^{\mu}\nabla_{\mu}$, the covariant D'Alembertian. Another possible form is:
\begin{eqnarray}
\partial_{\lambda}[\sqrt{-g}g^{\lambda\mu}]=0.
\label{harmong65775}
\end{eqnarray}

The harmonic gauge condition is suitable in the weak gravitational field limit when the metric can be expanded as $g_{\mu\nu}=\eta_{\mu\nu}+h_{\mu\nu}$ where $\eta_{\mu\nu}$ is the Minkowski metric and $h_{\mu\nu}$ is a small fluctuation.  In this gauge the Einstein field equations reduce in the weak limit to \cite{Carroll}:
\begin{eqnarray}
\Box h_{\mu\nu}-\frac{1}{2}\eta_{\mu\nu}\Box h=-16\pi G T_{\mu\nu},
\label{einst435}
\end{eqnarray}
where $h=h^{\mu}_{\mu}=h_{\mu\nu}\eta^{\mu\nu}$.

There are various gauge fixing conditions besides the harmonic gauge in general relativity.  Here we will mention only one another gauge related to the De Donder gauge \cite{Chen} and also analogous to the transverse gauge in electrodynamics. This is $g^{ij}\Gamma^{\rho}_{ij}=0$ where $i$ and $j$ refer to the spatial components of the metric. This gauge is very useful for studying gravitational fluxes and energies.

In this work we will consider a slight variation of an idea that was first proposed by Einstein in \cite{Einstein} in 1919 and further developed by Anderson and Finkelstein in \cite{Anderson} and discussed also in \cite{Finkelstein}-\cite{Smolin}. Namely we shall consider that the gauge fixing condition is given by $\Gamma^{\mu}_{\mu\lambda}=0$ which is equivalent to $\frac{\partial \ln \sqrt{|g|}}{\partial x^{\lambda}}=0$. It is clear then  that this gauge fixing condition leads to  a fixed measure of space time $\sqrt{-g}$ (unimodular gauge).  Then it makes sense that instead of discussing about a metric tensor one should introduce a conformal metric $f_{\mu\nu}$ which leads to concrete dynamical equations. One can then study the consequences of this gauge fixing choice for the general Lagrangian, for the curvature and  for the Einstein equations.

Here we will introduce a gauge fixing condition which is a slight variation of the unimodular gauge condition, namely  $\frac{\partial \ln \sqrt{|g|}}{\partial x^{\lambda}}=0$ (we will call this derivative gauge fixing).
In this gauge fixing the Ricci tensor will become:
\begin{eqnarray}
&&R_{ij}=\frac{\partial }{\partial x^j}\Gamma^j_{ik}-\frac{\partial }{\partial x^k} \Gamma^j_{ij}+
\nonumber\\
&&\Gamma^j_{js}\Gamma^s_{ik}-\Gamma^j_{ks}\Gamma^s_{ij}=
\nonumber\\
&&\frac{\partial }{\partial x^j}\Gamma^j_{ik}-\Gamma^j_{ks}\Gamma^s_{ij},
\label{ricic657}
\end{eqnarray}
so there is already a great simplification.  This simplification is more obvious for the scalar curvature:
\begin{eqnarray}
R=g^{ik}R_{ik}=g^{ik}\frac{\partial }{\partial x^j}\Gamma^j_{ik}-g^{ik}\Gamma^j_{ks}\Gamma^s_{ij}.
\label{curvature5454}
\end{eqnarray}
Let us analyze in more details the second term on the right hand side of Eq. (\ref{curvature5454}).

For a smooth manifold the covariant derivative of the metric tensor vanishes identically:
\begin{eqnarray}
\bigtriangledown_s g^{ij}=\partial_s g^{ij} +\Gamma^j_{sk}g^{ki}+\Gamma^i_{sk}g^{kj}=0.
\label{covmetr65774}
\end{eqnarray}
Then one may write:
\begin{eqnarray}
\Gamma^j_{sk}g^{ki}=-\partial_s g^{ij} -\Gamma^i_{sk}g^{kj},
\label{sommerel6776}
\end{eqnarray}
and furthermore,
\begin{eqnarray}
-g^{ik}\Gamma^j_{ks}\Gamma^s_{ij}=\partial_sg^{ij}\Gamma^s_{ij}+\Gamma^s_{ij}\Gamma^i_{sk}g^{kj}.
\label{res5464}
\end{eqnarray}
But,
\begin{eqnarray}
g^{kj}\Gamma^s_{ij}\Gamma^i_{sk}=\Gamma^s_{ij}\Gamma^j_{sk}g^{ki}.
\label{anothres645}
\end{eqnarray}
From Eqs. (\ref{res5464}) and (\ref{anothres645}) one then deduces:
\begin{eqnarray}
g^{ik}\Gamma^j_{ks}\Gamma^s_{ij}=-\frac{1}{2}\partial_sg^{ij}\Gamma^s_{ij}.
\label{finres54664}
\end{eqnarray}
We plug the result form Eq. (\ref{finres54664}) into Eq. (\ref{curvature5454}) to obtain in the gauge considered:
\begin{eqnarray}
R=g^{ik}\frac{\partial }{\partial  x^j}\Gamma^j_{ik}+\frac{1}{2}\Gamma^s_{ij}\partial_s g^{ij}.
\label{curvsecond657}
\end{eqnarray}
Since the determinant of the metric can be regarded  as constant one can consider that the total derivatives are zero, integrate by parts and  further write:
\begin{eqnarray}
&&R=\frac{1}{2}\frac{\partial}{\partial x^j}\Gamma^j_{ik}=
\nonumber\\
&&-\frac{\partial g^{ik}}{\partial X^m}\frac{\partial g_{mi}}{\partial x^k}+\frac{1}{2}\frac{\partial g^{ik}}{\partial x_m}\frac{\partial g_{ik}}{\partial x^m}.
\label{finres54675}
\end{eqnarray}
Thus the curvature becomes very amenable and with similarities with the gauge kinetic term for an abelian gauge field.

Einstein equations are then written with $R_{ij}$ from Eq. (\ref{ricic657}) and with $R$ from Eq. (\ref{finres54675}).  In the classical case the derivative  gauge fixing condition  can be introduced in the Lagrangian in the form of a vector Lagrangian multiplier as follows:
\begin{eqnarray}
&&S=\int d^4 x \sqrt{-g}\Bigg[\frac{1}{2k}R+{\cal L}_m \Bigg]-
\nonumber\\
&&\int d^4 x\Bigg[\lambda^{\mu}[\partial_{\mu}\ln\sqrt{-g}]\Bigg],
\label{intr5465}
\end{eqnarray}
where $\lambda^{\mu}$ are the Lagrange multipliers.

Note that in the case of unimodular gauge the constraint is introduced at the classical level through a single Lagrange multiplier $\lambda$ as in:
\begin{eqnarray}
&&S=\int d^4 x \sqrt{-g}\Bigg[\frac{1}{2k}{\cal R}+{\cal L}_m\Bigg] -
\nonumber\\
&&\int d^4 x\Bigg[\lambda[\sqrt{-g}-\epsilon]\Bigg],
\label{intr546545}
\end{eqnarray}
where $\epsilon$ is a constant. Then the variation with respect to the metric leads to the usual Einstein with a cosmological constant term (for a review see \cite{Bufalo} and the references therein):
\begin{eqnarray}
R_{\mu\nu}-\frac{1}{2}Rg_{\mu\nu}+\frac{k}{2}g_{\mu\nu}=\frac{k}{2}T_{\mu\nu}.
\label{unim87766}
\end{eqnarray}
It is evident from Eqs. (\ref{intr5465}) and  (\ref{intr546545}) that the derivative and unimodular gauge conditions are completely equivalent at the classical level up to a surface term.

Consequently  the unimodular gauge should be used as an alternative to the harmonic gauge for the case when the gravitational field is arbitrary and not necessarily weak.

The main reason for introducing  the gauge fixing condition instead as $\partial_{\mu}\ln\sqrt{-g}=0$ is that it permits a straightforward generalization to the quantum case by analogy with gauge fixing for quantum electrodynamics. First we update the derivative gauge fixing to $\partial_{\mu}\ln\sqrt{-g}=a_{\mu}$ where $a_{\mu}(x)$ is an arbitrary vector. Then in the quantum partition function besides all other factors that might appear there is a factor pertaining to the gauge condition:
\begin{eqnarray}
\int d g_{\mu\nu}\delta(\partial_{\mu}\ln\sqrt{-g}-a_{\mu})\exp[iS],
\label{qunatpartfunc6779}
\end{eqnarray}
which can be further written as (if gaussian weighting functions centered at zero are introduced):
\begin{eqnarray}
&&\int d a_{\mu}\exp[-i\int d^4x \sqrt{-g}\frac {a_{\mu}^2}{\epsilon^2}]\times
\nonumber\\
&&\int d g_{\mu\nu}\delta(\partial_{\mu}\ln\sqrt{-g}-a_{\mu})\exp[iS]=
\nonumber\\
&&\int d g_{\mu\nu}\exp[iS]\times
\nonumber\\
&&\exp[-i\int d^4 x \frac{1}{\epsilon}(\partial^{\mu}\ln\sqrt{-g})(\partial_{\mu}\ln\sqrt{-g})].
\label{finalres64554}
\end{eqnarray}

Next we shall discuss the implications of Eq. (\ref{finalres64554}) for the Einstein equations and for the particular limit of the weak gravitational field.

First the Einstein equations will become:
\begin{eqnarray}
&&R_{\mu\nu}-\frac{1}{2}g_{\mu\nu}R+
\nonumber\\
&&\frac{1}{2\epsilon}g_{\mu\nu}g^{ik}\frac{\partial \ln\sqrt{-g}}{\partial x^i}\frac{\partial \ln\sqrt{-g}}{\partial x^k}-
\nonumber\\
&&\frac{1}{\epsilon}g^{ik}\frac{\partial g_{\mu\nu}}{\partial x^k}\frac{\partial \ln\sqrt{-g}}{\partial x^i}=\frac{k}{2}T_{\mu\nu}.
\label{finaleisnted5454}
\end{eqnarray}


Next we shall consider the weak gravitational limit where $g_{\mu\nu}=\eta_{\mu\nu}+h_{\mu\nu}$. In this case the gauge fixing term in quadratic order becomes ($h=h_{\mu\nu}\eta_{\mu\nu}$):
\begin{eqnarray}
{\cal L}_{\epsilon}=-\frac{1}{4\epsilon}\eta^{ik}\frac{\partial h}{\partial x^i}\frac{\partial h}{\partial x^k}.
\label{gaugefix75664}
\end{eqnarray}
The gravity Lagrangian in the weak gravitational limit is \cite{Carroll}:
\begin{eqnarray}
&&{\cal L}=\frac{1}{2}\Bigg[\partial_{\mu}h^{\mu\nu}(\partial_{\nu}h)-(\partial_{\mu} h^{\rho\sigma}(\partial_{\rho}h^{\mu}_{\sigma})+
\nonumber\\
&&\frac{1}{2}\eta_{\mu\nu}(\partial_{\mu}h^{\rho\sigma})(\partial_{\nu}h_{\rho\sigma})-\frac{1}{2}\eta^{\mu\nu}(\partial_{\mu} h)(\partial_{\nu}h)\Bigg].
\label{res5435678}
\end{eqnarray}
Then the choice $\epsilon=-1$ in Eq. (\ref{gaugefix75664}) cancels the last term in Eq. (\ref{res5435678}).

In this work we discussed a gauge which is a slight modification of the unimodular gauge. Whereas the unimodular gauge can be very important through the presence of a cosmological constant term in practice its implementation requires the presence of a Lagrangian multiplier and also that of a surface term in the action which may lead to difficulties in the quantization of gravity \cite{Henneaux}. The harmonic gauge on the other hand is very amenable in the weak gravitational limit where it leads to drastic simplifications. Here we presented a more general version of the unimodular gauge (the derivative gauge) which has features that makes it attractive from two points of views. First it leads to some simplification in the weak gravitational limit not as good as the harmonic gauge butuseful. Then it is very convenient from the point of view of quantization because uncanning similarity with the Lorentz gauge for a gauge. We sketched a method through which the gauge fixing term might arrive in the Lagrangian without much details and discuss some consequences of this choice. More detailed derivations and analysis will be presented in a further work.

\end{document}